\documentclass[aps,prl,showpacs,twocolumn]{revtex4}
\usepackage{amsmath, amssymb}
\usepackage{graphicx}
\usepackage{subfigure}

\DeclareMathOperator{\erfi}{erfi}

\renewcommand{\vec}[1]{\mathbf{#1}}

\begin{document}

\title{Bell-type inequalities for arbitrary observables}

\author{E. Shchukin}
\email{evgeny.shchukin@gmail.com}
\author{W. Vogel}
\email{werner.vogel@uni-rostock.de}
\affiliation{Arbeitsgruppe Quantenoptik, Institut f\"{u}r  Physik,
Universit\"at Rostock, D-18051 Rostock, Germany}

\begin{abstract}
We derive a Bell-type inequality for observables with arbitrary spectra.  
For the case of continuous variable systems we propose a possible experimental
violation of this inequality, 
by using squeezed light and homodyne detection together with methods of
quantum-state reconstruction.  
It is shown that the violation is also possible for realistic detection
efficiencies.
\end{abstract}


\maketitle

The theory of Bell inequalities has a rather long history. The theory itself is
named after Bell's work 
\cite{ph-1-195}, where he proposed a quantitative approach to the old problem of
the incompleteness of 
Quantum Mechanics, posed in the famous EPR (Einstein, Podolsky, Rosen) paper
\cite{pr-47-777}. 
The first result 
of what is now called Bell-type inequality is the CHSH (Clauser, Horne, Shimony,
Holt) inequality \cite{prl-23-880}. 
For the first time experimental violations of Bell inequalities were observed in
\cite{prl-45-617}. Since that time 
there appeared a lot of other results, but the progress was mainly restricted to
finite-dimensional systems and 
observables with discrete spectra. 

The literature on Bell inequalities is quite extensive, both the theoretical and
experimental one, so here we 
restrict ourselves mainly to some important results which are relevant in the
context of our approach. The CHSH 
inequality was generalized to the so-called Mermin multiqubit inequalities
\cite{prl-65-1838}. The Mermin 
inequalities where further extended to a complete set of inequalities for an
arbitrary number of qubits \cite{pra-64-032112}. Mermin inequalities are distinguished for their quantum
violation, which is maximal in this 
complete set of inequalities and it exponentially depends on the number of
qubits. Among the recent trends, Bell 
inequalities for graph states \cite{prl-95-120405} and applications of Bell
states to secure quantum key 
distribution \cite{prl-97-120405} have been studied. 

It is well known that Bell inequalities are fulfilled by local-hidden variable theories and their violation is a clear quantum effect.
Recently an incompatibility theorem has been introduced~\cite{fp-33-1469},
which considers nonlocal-hidden variable theories. It has been  
experimentally demonstrated that Quantum Mechanics violates non-local realistic theories~\cite{n-446-871}, for further generalizations of the  incompatibility theorem, cf.~\cite{prl-99-210406}.

As already mentioned, most of the studies dealt with observables with discrete
spectra, usually dichotomic ones. 
For applications of the CHSH inequality for dichotomic variables to  continuous-variable quantum states see, for example, \cite{job-5-S333, pra-79-012101}. Only recently some results for observables
with arbitrary spectra have been 
obtained. The first multipartite Bell-type inequality for observables with
arbitrary spectra has been given in 
\cite{prl-99-210405}. In the bipartite case it reads as
\begin{equation}\label{eq:b}
\begin{split}
    \langle \hat{A}_1 &\hat{B}_1 - \hat{A}_2 \hat{B}_2 \rangle^2 + \langle
\hat{A}_1 \hat{B}_2 + \hat{A}_2 \hat{B}_1 \rangle^2 \\
    &\leqslant \langle \hat{A}^2_1 \hat{B}^2_1 \rangle + \langle \hat{A}^2_1
\hat{B}^2_2 \rangle + \langle \hat{A}^2_2 \hat{B}^2_1 \rangle + \langle
\hat{A}^2_2 \hat{B}^2_2 \rangle,
\end{split}
\end{equation}
where arbitrary Hermitian operators $\hat{A}_1$, $\hat{A}_2$ and $\hat{B}_1$,
$\hat{B}_2$ act on different sites. 
For two-qubit systems and dichotomic observables, the right-hand side of this
inequality reduces simply to $4$, so 
that it reads as
\begin{equation}\label{eq:b2}
    \langle \hat{A}_1 \hat{B}_1 - \hat{A}_2 \hat{B}_2 \rangle^2 + \langle
\hat{A}_1 \hat{B}_2 + \hat{A}_2 \hat{B}_1 \rangle^2 \leqslant 4. 
\end{equation}
It has been proved by Uffink \cite{prl-88-230406}, that any state (irrespective if it is separable or not)
of a two-qubit system satisfies 
this inequality. Hence the Bell-type inequality
\eqref{eq:b} can never be violated in this case. 
Moreover, for continuous-variable systems this inequality cannot be violated in
the case of quadrature measurements 
\cite{prl-101-040404}. Violations of the multipartite version of this inequality
for quadratures have been obtained 
in the original work \cite{prl-99-210405}, for the number of parties being at
least ten. This shows that violations 
of the inequality \eqref{eq:b} and its multipartite versions are not easy to
observe. 

In the present contribution we introduce another inequality which can be easily
violated both by the simplest two-quibit Bell 
state and by continuous-variable states. As an example, we simulate an
experiment with a two-mode squeezed vacuum. 
The violation can be demonstrated for realistic values of the squeezing parameter and of the detection
efficiencies.

Our main result is the following statement: any bipartite separable quantum
state satisfies the inequality
\begin{equation}\label{eq:bb}
\begin{split}
    \bigl\langle (&\hat{A}_1 \hat{B}_1 - \hat{A}_2 \hat{B}_2)^2 + (\hat{A}_1
\hat{B}_2 + \hat{A}_2 \hat{B}_1)^2 \bigr\rangle \\
    &\geqslant\langle\hat{A}_1\hat{B}_1\rangle^2 +
\langle\hat{A}_1\hat{B}_2\rangle^2 + \langle\hat{A}_2\hat{B}_1\rangle^2 +
\langle\hat{A}_2\hat{B}_2\rangle^2.
\end{split}
\end{equation}
At a first look it resembles the inequality \eqref{eq:b}. The difference is that
the squaring, averaging and the inequality sign are exchanged. The proof is based on the simple 
fact that any numbers $a$, $b$, $c$ and $d$ with $ad=bc$ satisfy the equality
\begin{equation}\label{eq:abcd}
    (a-d)^2 + (b+c)^2 = a^2 + b^2 + c^2 + d^2.
\end{equation}
Let us take Hermitian operators $\hat{A}_1$, $\hat{A}_2$, acting on one mode,
and $\hat{B}_1$, $\hat{B}_2$, acting 
on the other one, and  set $a = \langle\hat{A}_1\hat{B}_1\rangle$, $b =
\langle\hat{A}_1\hat{B}_2\rangle$, 
$c = \langle\hat{A}_2\hat{B}_1\rangle$, $d = \langle\hat{A}_2\hat{B}_2\rangle$.
Then for a factorizable state, 
$\hat{\varrho} = \hat{\varrho}^A \otimes \hat{\varrho}^B$, we have $ad = bc$ and
the identity \eqref{eq:abcd} gives 
us the relation
\begin{equation}\label{eq:f}
\begin{split}
    & \langle \hat{A}_1 \hat{B}_1 - \hat{A}_2 \hat{B}_2\rangle^2 + \langle
\hat{A}_1 \hat{B}_2 + \hat{A}_2 \hat{B}_1\rangle^2 \\
    &= \langle\hat{A}_1\hat{B}_1\rangle^2 + \langle\hat{A}_1\hat{B}_2\rangle^2 +
\langle\hat{A}_2\hat{B}_1\rangle^2 + \langle\hat{A}_2\hat{B}_2\rangle^2.
\end{split}
\end{equation}
From the non-negativity of the variance of Hermitian operators we get 
\begin{equation}
\begin{split}
    \langle (\hat{A}_1 \hat{B}_1 &- \hat{A}_2 \hat{B}_2)^2\rangle + \langle
(\hat{A}_1 \hat{B}_2 + \hat{A}_2 \hat{B}_1)^2\rangle \\
    &\geqslant \langle \hat{A}_1 \hat{B}_1 - \hat{A}_2 \hat{B}_2\rangle^2 +
\langle \hat{A}_1 \hat{B}_2 + \hat{A}_2 \hat{B}_1\rangle^2,
\end{split}
\end{equation}
which, by inserting Eq.~\eqref{eq:f}, proves the correctness of the inequality
\eqref{eq:bb} for factorizable states. 

To extend this inequality to arbitrary separable states, we use the
Cauchy-Schwarz inequality $( \sum_k p_k x_k )^2 
\leqslant \sum_k p_k x^2_k$, where $\{p_k\}$ is a probability distribution and
$x_k$ are arbitrary real numbers. 
This inequality expresses the non-negativity of the variance of a random
variable. Let us consider a separable 
state $\hat{\varrho} = \sum_k p_k \hat{\varrho}^A_k \otimes \hat{\varrho}^B_k
\equiv \sum_k p_k \hat{\varrho}_k$ 
and estimate the right hand sideboth of the inequality \eqref{eq:bb}. We have
\begin{equation}
\begin{split}
    &\langle\hat{A}_1\hat{B}_1\rangle^2_{\hat{\varrho}} +
\langle\hat{A}_1\hat{B}_2\rangle^2_{\hat{\varrho}} +
\langle\hat{A}_2\hat{B}_1\rangle^2_{\hat{\varrho}} +
\langle\hat{A}_2\hat{B}_2\rangle^2_{\hat{\varrho}} \\ 
    &\leqslant \sum_k p_k
\bigl(\langle\hat{A}_1\hat{B}_1\rangle^2_{\hat{\varrho}_k} + \ldots +
\langle\hat{A}_2\hat{B}_2\rangle^2_{\hat{\varrho}_k}\bigr) \\
    &\leqslant \sum_k p_k \bigl\langle (\hat{A}_1 \hat{B}_1 - \hat{A}_2
\hat{B}_2)^2 + (\hat{A}_1 \hat{B}_2 + \hat{A}_2 \hat{B}_1)^2
\bigr\rangle_{\hat{\varrho}_k} \\
    &= \bigl\langle (\hat{A}_1 \hat{B}_1 - \hat{A}_2 \hat{B}_2)^2 + (\hat{A}_1
\hat{B}_2 + \hat{A}_2 \hat{B}_1)^2 \bigr\rangle_{\hat{\varrho}}.
\end{split} 
\end{equation}
The first step is due to the Cauchy-Schwarz inequality and the second one is due
to the fact that the inequality 
\eqref{eq:bb} is valid for factorizable states, as it has already been proved.
The last step simply expresses the 
linearity of the mean value, which completes the proof.

If we set $a = a_1 b_1$, $b = a_1 b_2$, $c = a_2 b_1$, $d = a_2 b_2$, then $ad =
bc$ and the identity 
\eqref{eq:abcd} simply expresses the multiplicativity of the norm of complex
numbers
\begin{equation}
    |a_1 + i b_1|^2 |a_2 + i b_2|^2 = |(a_1 + i b_1)(a_2 + i b_2)|^2.
\end{equation}
We can write this equality for any number of factors and get corresponding
multipartite Bell-type inequalities. 
We can go further and get Bell-type inequalities with more observables. Norms of
the algebras of quaternions and 
octonions are multiplicative, and this property can be used to obtain the square
identities generalizing the one 
given by Eq.~\eqref{eq:abcd}, which can be used to derive multipartite Bell-type
inequalities with four and eight 
observables per site. Details of this approach have been given in
\cite{pra-78-032104}. In this contribution we will 
concentrate on multipartite inequalities with two observables per site.

After simple algebraic manipulations the inequality \eqref{eq:bb} can be
rewritten in the form
\begin{equation}\label{eq:sigma}
    \sigma^2_{\hat{A}_1\hat{B}_1} + \sigma^2_{\hat{A}_1\hat{B}_2} +
\sigma^2_{\hat{A}_2\hat{B}_1} + \sigma^2_{\hat{A}_2\hat{B}_2} \geqslant |\langle
[\hat{A}_1, \hat{A}_2] [\hat{B}_1, \hat{B}_2] \rangle|,
\end{equation}
where $\sigma^2_{\hat{A}} = \langle \hat{A}^2 \rangle - \langle \hat{A}
\rangle^2$ is the square of the variance of 
the operator $\hat{A}$. The approach of \cite{prl-99-210405} to derive the
inequality \eqref{eq:b} is to ignore 
local commutators, which would give a trivial result in our case. We see that
separability puts a more strict 
condition on the sum of the variances then the one expressed by the inequality
\eqref{eq:b3}. The product of two 
commutators on the right hand side is a product of two local observables since
it can be represented as 
$[\hat{A}_1, \hat{A}_2] [\hat{B}_1, \hat{B}_2] = -i[\hat{A}_1, \hat{A}_2]
i[\hat{B}_1, \hat{B}_2]$. We will show 
that the inequality \eqref{eq:sigma} can be easily violated. For the strength of
violation, $V$, we use the ratio 
of the right hand side (containing commutators) and the left hand side (the sum
of squares of dispersions). When 
this ratio exceeds one, $V\ge 1$, then the inequality \eqref{eq:sigma} is
violated. In such a case the maximal 
violation for a given state is the maximum of this ratio for all possible
choices of the operators $\hat{A}_k$ and 
$\hat{B}_k$. The inequality \eqref{eq:b3} shows that the maximal violation
cannot exceed $2$. Below we will see 
that this limit can be easily achieved.

It is interesting to note that an arbitrary state of a two qubit system satisfies the
following inequality:
\begin{equation}\label{eq:b3}
    \sigma^2_{\hat{A}_1\hat{B}_1} + \sigma^2_{\hat{A}_1\hat{B}_2} +
\sigma^2_{\hat{A}_2\hat{B}_1} + \sigma^2_{\hat{A}_2\hat{B}_2} \geqslant
\frac{1}{2}|\langle [\hat{A}_1, \hat{A}_2] [\hat{B}_1, \hat{B}_2] \rangle|.
\end{equation}
It differs from the inequality \eqref{eq:sigma} only by a constant factor of $1/2$ on the 
right-hand side. The inequality \eqref{eq:b3} can be obtained from \eqref{eq:b2} using methods of 
\cite{prl-88-230406}. To illustrate violations of the inequality \eqref{eq:sigma}, consider a
two-qubit system. Let us take the operators 
$\hat{A}_k = \hat{\sigma}^A_{\vec{r}_k}$, $\hat{B}_k =
\hat{\sigma}^B_{\vec{s}_k}$, where the normalized vectors 
$\vec{r}_k$ and $\vec{s}_k$ represent the direction along which the spin
projections are measured. Then the 
inequality \eqref{eq:sigma} becomes
\begin{equation}\label{eq:bs}
\begin{split}
    4-\langle \hat{\sigma}^A_{\vec{r}_1} \hat{\sigma}^B_{\vec{s}_1} \rangle^2 &-
\langle \hat{\sigma}^A_{\vec{r}_1} \hat{\sigma}^B_{\vec{s}_2} \rangle^2 -
\langle \hat{\sigma}^A_{\vec{r}_2} \hat{\sigma}^B_{\vec{s}_1} \rangle^2 -
\langle \hat{\sigma}^A_{\vec{r}_2} \hat{\sigma}^B_{\vec{s}_2} \rangle^2 \\
    &\geqslant 4|\langle\hat{\sigma}^A_{\vec{r}_1\times\vec{r}_2}
\hat{\sigma}^B_{\vec{s}_1\times\vec{s}_2}\rangle|.
\end{split}
\end{equation}
For the Bell state $|\Phi\rangle = (1/\sqrt{2})(|01\rangle + |10\rangle)$ we
have $\langle \hat{\sigma}^A_{\vec{r}} 
\hat{\sigma}^B_{\vec{s}} \rangle = r_x s_x + r_y s_y - r_z s_z =
(\tilde{\vec{r}}, \vec{s})$, where we set 
$\tilde{\vec{r}} = (r_x, r_y, -r_z)$; $\tilde{\vec{r}}$ is the reflection of
$\vec{r}$ with respect to the $xy$-plane. Note that $\widetilde{\vec{r}_1 \times
\vec{r}_2} = - \tilde{\vec{r}}_1 \times \tilde{\vec{r}}_2$. For the vectors
$\tilde{\vec{r}}_1$, $\tilde{\vec{r}}_2$, $\vec{s}_1$ and $\vec{s}_2$ lying in
the same plane with the straight angles between $\tilde{\vec{r}}_1$,
$\tilde{\vec{r}}_1$ and $\vec{s}_1$, $\vec{s}_2$ the inequality \eqref{eq:b3} is
satisfied with the equality sign. This implies that the Bell-type inequality \eqref{eq:sigma} is 
violated by a factor of $2$.

For continuous variable states let us introduce the analogues of the Pauli
operators via
\begin{equation}
\begin{split}
    \hat{S}_{X, N} &= \sum^N_{k=0}(|2k\rangle\langle 2k+1| + |2k+1\rangle\langle
2k|), \\
    \hat{S}_{Y, N} &= -i\sum^N_{k=0}(|2k\rangle\langle 2k+1| -
|2k+1\rangle\langle 2k|), \\
    \hat{S}_{Z, N} &= \sum^{2N+1}_{k=0}(-1)^k|k\rangle\langle k|, \ \hat{S}_{0,
N} = \sum^{2N+1}_{k=0}|k\rangle\langle k|.
\end{split}
\end{equation}
The first of these operators commutes with the other three, which have the
commutator rules 
$[\hat{S}_{\alpha, N}, \hat{S}_{\beta, N}] = 2i\hat{S}_\gamma$,  $\{\alpha,
\beta, \gamma\} = \{X, Y, Z\}$, of the 
ordinary Pauli operators, though their squares $\hat{S}^2_{X, N} = \hat{S}^2_{Y,
N} = \hat{S}^2_{Z, N} = 
\hat{S}_{0, N}$ are not the unity operator unless $N = +\infty$. In this
limiting case we denote them simply as 
$\hat{S}_X$, $\hat{S}_Y$ and $\hat{S}_Z$. The operators $\hat{S}_{\vec{r}, N}$
and $\hat{S}_{\vec{r}}$ are defined 
in full analogy with $\hat{\sigma}_{\vec{r}}$. 

\begin{figure}
    \includegraphics{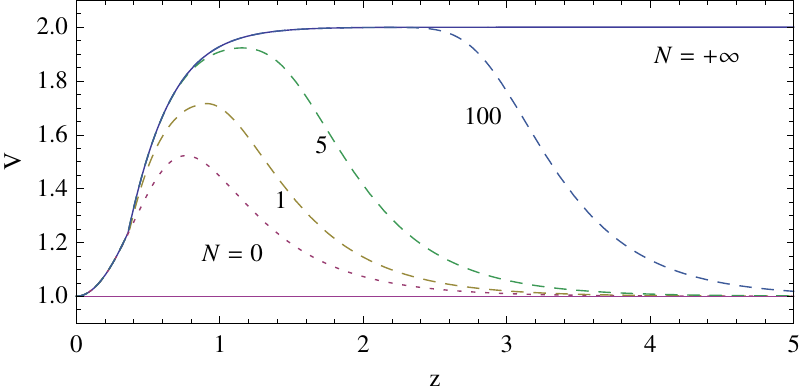}
\caption{(Color online). Maximal violation of the inequality \eqref{eq:sigma}
for the squeezed vacuum for different 
values of $N$.} \label{fig:fig1}
\end{figure}

As a realistic example of a continuous variable state, let us consider the
two-mode squeezed vacuum given by 
$|\Psi\rangle = S(z) |00\rangle$, where the two-mode squeezing operator is
defined as $S(z) = \exp(z^* \hat{a} 
\hat{b} - z \hat{a}^\dagger \hat{b}^\dagger)$. For this state (for real
squeezing parameter $z$) we have
\begin{equation}
\begin{split}
    \langle \hat{S}^A_{\vec{r}, N} &\hat{S}^B_{\vec{s}, N} \rangle =
(1-\tanh^{4(N+1)}(z)) \times \\
    &(-\tanh(2z) r_x s_x + \tanh(2z) r_y s_y + r_z s_z)
\end{split}
\end{equation}
and $\langle \hat{S}^{A2}_{\vec{r}, N} \hat{S}^{B2}_{\vec{s}, N} \rangle =
1-\tanh^{4(N+1)}(z)$. Figure 
\ref{fig:fig1} shows the numerically calculated maximal violation of the
inequality \eqref{eq:sigma} for some finite 
values of $N$. The solid line corresponds to the case of $N = +\infty$. Note
that for $N = +\infty$ we simply have 
\begin{equation}
    \langle \hat{S}^A_{\vec{r}, N} \hat{S}^B_{\vec{s}, N} \rangle = -\tanh(2z)
r_x s_x + \tanh(2z) r_y s_y + r_z s_z, 
\end{equation}
which in the limit $z \to +\infty$ coincides with
$\langle\hat{\sigma}^A_{\vec{r}}\hat{\sigma}^B_{\vec{s}}\rangle$ 
for the Bell state after reordering the components of $\vec{r}$ and $\vec{s}$.
Thus, in this case the violation tends 
to $2$ when $z \to +\infty$.

Let us now discuss experimental applications of our Bell-type inequality. Matrix
elements $\varrho_{nm} = 
\langle n|\hat{\varrho}|m\rangle$ of a density operator $\hat{\varrho}$ can be
obtained from the quadrature 
distribution $p_\varphi(x) = {}_\varphi\langle
x|\hat{\varrho}|x\rangle_\varphi$, where $|x\rangle_\varphi$ is the 
eigenstate of the quadrature operator $\hat{x}_\varphi |x\rangle_\varphi = x
|x\rangle_\varphi$,  according to the 
reconstruction formula
\begin{equation}
    \varrho_{nm} = \frac{1}{\pi}\int^{\pi}_0 e^{i (n-m) \varphi}
\int^{+\infty}_{-\infty} p_\varphi(x) f_{nm}(x)\,dx\,d\varphi.
\end{equation}
The reconstruction kernels $f_{nm}(x)$ are given by $f_{nm}(x) = (\psi_n(x)
\varphi_m(x))'$ for $n \leqslant m$ and 
$f_{nm}(x) = f_{mn}(x)$ for $n>m$, where $\psi_n(x) = \langle x|n\rangle$ is the
well known wave function of the 
$n$th Fock state in the coordinate representation, the normalizable solution of
the Schr\"{o}dinger equation for the 
harmonic oscillator. Expressions for the non-normalizable solution
$\varphi_m(x)$ are also known. In 
\cite{oc-127-144} it is given as an action of a differential operator on the
function $\varphi_0(x)$ and in 
\cite{pra-61-063819} it is given explicitly in terms of the degenerate
hypergeometric function. Here we could derive 
the explicit expression
\begin{equation}\label{eq:varphi}
\begin{split}
    \varphi_m(x) &= \pi\psi_m(x)\erfi(x) -
\frac{2\sqrt[4]{\pi}e^{x^2/2}}{\sqrt{2^mm!}} \times\\
&\sum^{\left\lfloor\frac{m-1}{2}\right\rfloor}_{k=0}(-2)^kk!\binom{m-k-1}{k}H_{
m-2k-1}(x).
\end{split}
\end{equation}
Here $\lfloor x \rfloor$ is the floor of $x$, i.e. the largest integer that is
not greater than $x$.

%End

Up to now we have discussed only single mode case, but in the case of several
modes the reconstruction kernel is the product of the single mode kernels of the
corresponding modes. In the bipartite case under study, we have the two mode
quadrature distribution defined via $p_{\varphi\theta}(x, y) = {}_\varphi\langle
x|{}_\theta\langle y|\hat{\varrho}|x\rangle_\varphi|y\rangle_\theta$. The
average value $\langle \hat{S}^A_{\mu, N} \hat{S}^B_{\nu, N} \rangle$, where
$\mu, \nu = X, Y, Z, 0$, is given by
\begin{equation}\label{eq:me}
    \langle \hat{S}^A_{\mu, N} \hat{S}^B_{\nu, N} \rangle =
\int^{\pi}_0\int^{\pi}_0  \tilde{p}_{\mu\nu}(\varphi, \theta)
\Phi_\mu(\varphi) \Phi_\nu(\theta) \,d\varphi\,d\theta,
\end{equation}
where we denoted
\begin{equation}
    \tilde{p}_{\mu\nu}(\varphi, \theta) = \int^{+\infty}_{-\infty}
\int^{+\infty}_{-\infty} p_{\varphi\theta}(x, y) F_{\mu, N}(x) F_{\nu,
N}(y)\,dx\,dy.
\end{equation}
The phase dependent functions are $\Phi_X(\varphi) = \frac{2}{\pi}\cos\varphi$,
$\Phi_Y(\varphi) = \frac{2}{\pi}\sin\varphi$, $\Phi_Z(\varphi) = \Phi_0(\varphi)
= \frac{1}{\pi}$ and the reconstruction kernels are defined via
\begin{equation}
\begin{split}
    F_{X, N}(x) &= F_{Y, N}(x) = \sum^N_{k=0} f_{2k, 2k+1}(x), \\ 
    F_{\alpha, N}(x) &= \sum^{2N+1}_{k=0} u^k_\alpha f_{k, k}(x),
\end{split}
\end{equation}
where $u_0 = 1$ and $u_Z = -1$. In an experiment one obtains data in the form of
the table $\{x_{ij,k}, y_{ij,k}\}$, $i,j = 1, \ldots, N_{\mathrm{ph}}$, $k = 1,
\ldots N_{\mathrm{pos}}$, where one measures $N_{\mathrm{qu}}$ quadrature values
for each of $N^2_{\mathrm{ph}}$ chosen phase pairs $(\varphi_i, \theta_j)$. The
inner integrals in Eq.~\eqref{eq:me}, denoted as $\tilde{p}_{\mu\nu}(\varphi,
\theta)$, can be obtained by the sampling 
\begin{equation}\label{eq:pmn}
    \tilde{p}_{\mu\nu}(\varphi_i, \theta_j) =
\frac{1}{N_{\mathrm{qu}}}\sum^{N_{\mathrm{qu}}}_{k=1} F_{\mu, N}(x_{ij,k})
F_{\nu, N}(y_{ij,k}), 
\end{equation} 
with $i,j = 1, \ldots, N_{\mathrm{ph}}$. Having these numbers we can calculate
the outer integrals in Eq.~\eqref{eq:me} using the fast Fourier transform.

Recently, a quantum noise redunction of a factor of $10$ has been achieved
\cite{prl-100-033602}. Using this, we can estimate the degree of squeezing that
can be obtained in a realistic experiment. For the squeezed vacuum for the
dispersion of $\hat{x} = \hat{x}_A + \hat{x}_B$ we have $\langle(\Delta
\hat{x})^2\rangle = e^{-2z}$. To find the maximal squeezing parameter $z$ we
have to solve the equation $e^{-2z} = 0.1$, which gives $z \simeq 1.15$. For
example, for $N=5$ we get a violation of $\simeq 1.9$, which is very close to
the maximal possible value.

\begin{figure}
    \subfigure{\includegraphics[scale=0.7]{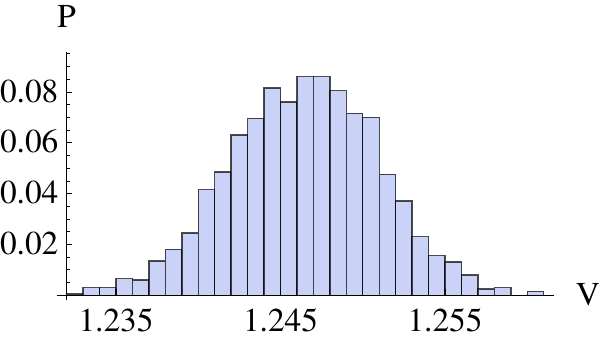}}
    \subfigure{\includegraphics[scale=0.7]{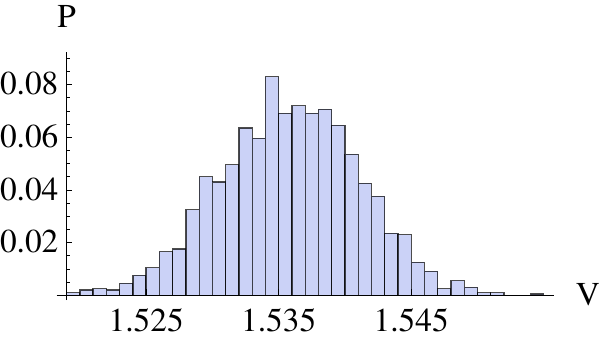}}
\caption{(Color online) Statistical distribution of the maximal violation obtained in a simulation of an
experiment with the detection efficiencies $\eta=0.9$ (left) and $\eta=1$ (right).} \label{fig:fig2a}
\end{figure}

Finally, we discuss the maximal violation of the inequality \eqref{eq:sigma}
with imperfect photodetector with the efficiency $\eta$. The moments of the
creation and annihilation operators $\mu_{n, n+p}(\eta) = \langle\hat{a}^n
\hat{a}^{\dagger n+p}\rangle_{\mathrm{meas}}$ measured by such a photodetector
are
$\mu_{n, n+p}(\eta) = \eta^{n+p/2} \mu_{n, n+p}$, where the moments on the right
hand side is the ``true'' moments, measured by the perfect photodetector with
the ideal efficiency $\eta=1$ \cite{jmo-34-709}. The density matrix elements can be
obtained from the moments according to the following expression
\cite{pra-46-6097}:
\begin{equation}\label{eq:re}
    \varrho_{n, m}(\eta) = \frac{1}{\sqrt{n!m!}} \sum^{+\infty}_{i=0}
\frac{(-1)^i}{i!} \mu_{n+i, m+i}(\eta).
\end{equation}
The extension to the multimode case is straightforward. Thus, we can calculate
the average values $\langle\hat{S}^A_{\vec{r}, 0}\hat{S}^B_{\vec{s}, 0}\rangle$
with the matrix elements obtained from the measurements with non-ideal
efficiency $\eta$. 

We have simulated an experiment by generating $N_{\mathrm{qu}} =
10^3$ points $(x_{ij,k}, y_{ij,k})$, $k = 1, \ldots, 10^3$ with the distribution
$p_{\varphi_i\theta_j}(x, y)$ for $i,j = 1, \ldots, N_{\mathrm{ph}}=61$, where
$\varphi_i = (\pi/60)(i-1)$, $\theta_j = (\pi/60)(j-1)$. These points were used to
calculate the quantities $\tilde{p}_{\mu\nu}(\varphi_i, \theta_j)$, $\mu,\nu =
X, Y, Z, 0$, according to Eq.~\eqref{eq:pmn}. Then we have used the simplest
numerical integration scheme to get the quantum mean values $\langle
\hat{S}^A_{\mu, N} \hat{S}^B_{\nu, N} \rangle$ according to Eq.~\eqref{eq:me}.
As an example we took $z=0.8$ and $\eta=0.9, 1.0$. Fig.~\ref{fig:fig2a} shows the 
distribution of the maximal violation $V$ for $2000$ runs of the simulation. One can 
calculate the maximal violation by using the
analytical expressions for $\langle \hat{S}^A_{\mu, N} \hat{S}^B_{\nu, N}
\rangle$ obtained with  Eq.~\eqref{eq:re}. Comparing with the values averaged
over $2000$ runs of the experiment simulations, we found that in all four cases
the difference is $\simeq 1$\%. 

In conclusion, we have derived a Bell-type inequality for arbitrary observables,
which can be violated for discrete as well as for continuous-variable
quantum states. We have simulated a realistic experimental violation of the inequality. For this purpose a two-mode squeezed-vacuum state has been considered. The methods of reconstructing the needed correlation functions are provided.

\end{document}